\documentclass{ws-mpla_tt}

\begin{document}

\markboth{Dieter L\"ust and Tomasz R. Taylor}
{Limits on stringy signals at the LHC}

\catchline{}{}{}{}{}

\title{LIMITS ON STRINGY SIGNALS AT THE LHC}

\author{\footnotesize DIETER L\"UST}

\address{Max--Planck--Institut f\"ur Physik, \\
 Werner--Heisenberg--Institut,
80805 M\"unchen, Germany\vspace{-3mm}}
\address{Arnold Sommerfeld Center for Theoretical Physics
Ludwig-Maximilians-Universit\"at M\"unchen,
80333 M\"unchen, Germany}

\author{TOMASZ R.\ TAYLOR}

\address{Department of Physics,\\
 Northeastern University, Boston, MA 02115, USA
}

\maketitle


\begin{abstract} We explain what features of string theory can be tested at the LHC.
We review the present bounds on the string mass and on extra gauge bosons as well as the prospects for
the upcoming experiments.
\end{abstract}


\section{Introduction}	
After the discovery of the Higgs-like particle with mass around 125 GeV by the ATLAS and CMS collaborations at CERN,
 still many convincing reasons exist to believe that the resolution of the
hierarchy problem lies in new
physics around the TeV mass scale. In fact, there are at least two, not necessarily mutually
exclusive scenarios, offered as solutions of the hierarchy problem:
\begin{itemlist}
\item Low energy supersymmetry at around 1 TeV.
\item Large extra dimensions and a low scale
for quantum gravity  at around 1 TeV \cite{ArkaniHamed:1998rs,Antoniadis:1998ig}.
\end{itemlist}
Here we discuss some rather generic features about string and brane compactifications, which are relevant for the search of new physics at the LHC and which might
provide concrete and fundamental realizations of the general scenarios mentioned above.
In fact, the relevance of strings for the experiments at the LHC is closely linked to the question what is the typical scale of string theory, called the string mass $M$.
A priori, considering D-brane compactifications with open strings on branes, the string scale is a free parameter, not being determined by the Standard Model
gauge interactions. Hence in this class of models $M$ can be in principle everywhere between the weak scale and the Planck scale. Therefore we divide
the discussion into two cases, namely into low and high string scale compactifications.

\subsection{Low string scale compactifications}
The most spectacular and most direct LHC signatures from strings will emerge in case $M$ is
very low, namely at the
order of TeV.
Then the gravitational and gauge interactions are unified at around 1 TeV,
and the observed weakness of gravity at lower energies is due to the existence
of large extra dimensions, where the volume $V$ of the internal space in units of $M$ must be of the order $VM^6={\cal O}(10^{32})$.
This scenario can be nicely realized by D-brane compactifications (for a review see e.g.\ [\refcite{Blumenhagen:2006ci}]). Here
gravitons, i.e.\ closed strings may scatter into the extra space and
by this the gravitational coupling constant is decreased to its observed value.
On the other hand, the particles of the Standard Model correspond to the zero modes of open strings that are confined to lower dimensional D-branes, which typically  fill
part of the extra-dimensions and also intersect among each other in the higher dimensional space.
A typical brane configuration\cite{Cremades:2003qj} is shown in Fig.1.\cite{Anchordoqui:2012wt}
\begin{figure}[h]
\centerline{\psfig{file=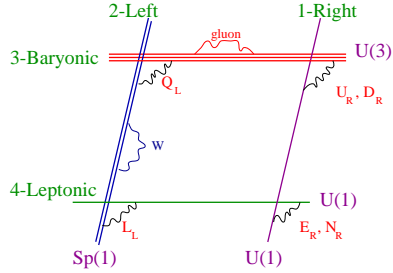,width=2.0in}}
\vspace*{8pt}
\caption{A typical configuration of intersecting branes, in this case corresponding to the so-called SM$^{++}$ model  with the gauge group $SU(3)\times SU(2)\times U(1)_Y\times U(1)^2$.}
\end{figure}
Gauge bosons emerge from open strings ending on a the same stack of overlapping branes, with U(1) gauge bosons ending on the same brane, while non-abelian gauge bosons like gluons stretch between distinct branes in a single stack. Quarks and leptons emerge from strings stretching between intersecting stacks of branes.
Moreover a whole tower of massive open  string excitations
will open up at around
$1\ {\rm TeV}$, where the new particles follow the well known Regge
``trajectories'' of vibrating strings,
\begin{equation}
M_n^2=n M^2\, ,\quad j=j_0+\alpha'  M_n^2\ ,
\end{equation}
with spin $j$ particles populating lines with the universal slope $\alpha'$ determined by the fundamental
string  mass scale $M^2=\alpha'^{-1}$, and ``intercepts'' $j_0$. At the $n$-th level, the spins of excited gluons range from 0 to $n+1$ and of excited quarks from 1/2 to $n+1/2$. The standard model spectrum is replicated at each level $n$ by  higher spin particles.

Let us list what kind of string signatures from a low string scale and from
large
extra dimensions in brane models can be
possibly expected  at the LHC:

\vskip0.3cm

\noindent (i) The discovery of string Regge excitations with masses of
order $M$. As we will explain in the following, the production cross sections and the decay rates of the heavy string excitations can be computed in a model
independent way for a large class of D-brane models, such that universal predictions for the discovery of these particles can be made.
In fact,
the production of string Regge excitations  will lead to new contributions to standard model
 scattering
processes, like QCD jets or scattering of quarks
into leptons or gauge bosons, which can be measurable at LHC in case the
string scale is low \cite{Cullen:2000ef,Chialva:2005gt,Lust:2008qc,Anchordoqui:2008di,Anchordoqui:2009mm,Lust:2009pz,Dong:2010jt}.

\vskip0.3cm

\noindent (ii) The discovery of new exotic particles around $M$.
Another very interesting signal for new stringy physics at the LHC is the production
of heavy neutral $Z'$ gauge bosons (see e.g.\ [\refcite{Kiritsis:2002aj,Ghilencea:2002da,Anchordoqui:2011ag,Anchordoqui:2011eg}]). These particles are quite generic in any string compactification,
and they receive their mass via a Green-Schwarz mixing with axionic scalar fields.
E.g.
in a four stack D6-brane model with gauge group $SU(3)\times SU(2)\times U(1)_Y\times U(1)^2$, see Fig.1, at least
one extra $U(1)$ gauge bosons will get a mass of order $g'M<M$ by the Green-Schwarz effect.

\vskip0.3cm

\noindent (iii) The discovery of (non-perturbative) quantum gravity effects in the form of
mini black
holes.
One of the most exciting possibilities for the LHC is the discovery of small
higher-dimensional black holes that can be
formed when two sufficiently energetic particles collide
\cite{Banks:1999gd,Giddings:2001bu,Dimopoulos:2001hw,Meade:2007sz}.
This means that effects of higher dimensional quantum gravity can get  strong if the string scale
is low around the TeV scale, and if the volume of the extra dimensions is large.
The geometrical cross section for the production of mini black holes is of the order
\begin{equation}
\sigma(E)\sim{1\over M_{b.h.}^2}\Biggl({E\over M_{b.h.}}\Biggr)^\alpha\, ,
\end{equation}
where $M_{b.h.}$ is the black hole mass, i.e. the effective scale of quantum gravity, and $\alpha\leq 1$
for higher dimensional black holes. Since the production of mini black holes is basically
a non-perturbative effect, the black hole mass is suppressed by the string
coupling constant compared to the string scale:
\begin{equation}
M_{b.h.}\sim{M\over g_{\rm string}}\,.
\end{equation}
Therefore, for weak string coupling, the threshold for black hole production
is almost inevitably higher
than the threshold for the production of string resonances.
\vskip0.3cm

\noindent (iv) The discovery of Kaluza-Klein (KK) excitations. In superstring theory with D-branes there are two classes of Kaluza-Klein states: bulk excitations of closed string states -- like KK graviton etc.\ -- and the excitations associated to the extra dimensions of D-branes extending into the bulk space. The masses of bulk states are determined by the properties of the six-dimensional Calabi-Yau manifold. The KK excitations of D-brane states are charged under the respective gauge group and their masses depend on how D-branes wrap into closed surfaces (cycles) inside Calabi-Yau manifold. Experimental signatures of KK states are common to all extra-dimensional extensions of the standard model and will be discussed elsewhere in this volume.

\subsection{High string scale compactifications}

Now let us turn to string compactifications with the string scale much above the TeV region. In fact,
there are no compelling reasons why the string mass scale should be much lower than the
Planck mass.
In the large volume compactifications of
Refs.[\refcite{Balasubramanian:2005zx,Conlon:2005ki,Conlon:2007xv}]
it was shown that that one can  stabilize
moduli in such a way that the string scale $M$
is at intermediate energies of about $10^{11-12}~{\rm GeV}$. Then
the internal CY volume $V$ is of order $VM^6={\cal O}(10^{16})$. The motivation for
this scenario is to obtain a low supersymmetry breaking scale around $1~{\rm TeV}$, since
one derives the following relation for the gravitino mass:
\begin{equation}
m_{3/2}\sim {M^2\over M_{\rm Planck}}\, .
\end{equation}
Hence the experimental indications for this kind of scenario at the LHC would be given by usual signatures for low energy supersymmetry, whose details largely
depend on further assumptions of the concrete compactification scheme.

Finally the string scale can be also around the Planck scale or also a bit lower around the GUT scale of about $10^{16}$ GeV, a choice, which is motivated
by the unification of the gauge coupling constants of the Standard Model. However D-brane models with such a high string scale can nevertheless lead to some rather model
independent new physics predictions around the TeV scale. Specifically, as discussed in [\refcite{Anchordoqui:2012wt}], in generic 4-stack D-brane models
one typically arrives at a simple extension of the Standard Model, called SM$^{++}$ (see Fig.1) obtained by adding to the scalar
  sector a complex $SU(2)$ singlet that has the quantum numbers of the
  right-handed neutrino, $H''$, and to the gauge sector an $U(1)$ that
  is broken by the vacuum expectation value of $H''$ (see also [\refcite{iso}]).  The associated
$Z''$-gauge boson can be relatively light.

\section{Low string scale: string resonances in dijets}
The lightest string resonances are excited quarks and gluons which should be copiously produced at the LHC energies just above the string threshold. They are produced by gluon-gluon or quark-gluon fusion, at the rates typical for QCD processes. These resonances decay into gluon pairs or quarks and gluons, giving rise to spectacular resonance bumps in dijet cross sections. As an example, consider the lightest quark resonances $q^*$, which appear in two separate spin $j=1/2$ and $j=3/2$ eigenstates. They give rise to a peak in the squared modulus of the quark-gluon scattering amplitude, at the center of mass energies close to their mass $M$:
\begin{eqnarray}
|{\cal M}(qg \to qg)|^2  ~=~ - \frac{4}{9} \frac{g^4}{M^2}\
&& \!\!\!\left[  \frac{M^4 \hat  u}{( \hat s-M^2)^2 + (\Gamma_{q^*}^{j=1/2} M)^2}\right.\nonumber\\ && +\left. \frac{\hat u^3}{(\hat s-M^2)^2 + (\Gamma_{q^*}^{j=3/2} M)^2}\right],
\label{qg}
\end{eqnarray}
where $\hat s$ and $\hat u$ are the Mandelstam invariants of this particular subprocess, and $g$ is the QCD coupling constant $(\alpha_{\rm QCD}=\frac{g^2}{4\pi}\approx 0.1)$. Here, the resonance poles have been softened to a Breit-Wigner form with the partial widths
$\Gamma_{q^*}^{j=1/2} = \Gamma_{q^*}^{j=3/2} = 37\, (M/{\rm
  TeV})~{\rm GeV}$,
as calculated in Ref.[\refcite{Anchordoqui:2008hi}].
The squared amplitudes for all $2\to 2$ and $2\to 3$ parton subprocesses are collected in Refs.[\refcite{Lust:2008qc},\refcite{Lust:2009pz}] while the decay rates can be found in Ref.[\refcite{Anchordoqui:2008hi}]. It is remarkable that the amplitude (\ref{qg}), as well as the amplitude describing $gg\to gg$ scattering, do {\em not\/} contain any reference to extra dimensions. Recall that one of the most common criticisms of string theory is that it does not predict the shape or size of extra dimensions responsible for the properties of low energy effective theory -- usually called the landscape problem --  but here, this problem has been circumvented by a complete {\em universality\/} of the amplitudes. As long as quark and gluons appear in the low energy spectrum, string resonances contribute in a universal way, common to {\em all} compactifications. There are conceivably no other extensions of the standard model that would produce such unambiguous, spectacular signatures of strongly interacting, higher spin particles. In reality, string resonances are very easy to discover.

The amplitudes involving four and more fermions, like $q\bar q\to q\bar q$, do depend on the details of compactification, however they do not significantly contribute  to dijet cross sections for two reasons: 1) color triplets yield smaller  $SU(3)$ QCD factors than gluon octets 2) at high energies (large Feynman's $x$), antiquark lumnosities inside protons are lower than quark and  gluon luminosities. Actual dijet mass distributions are obtained by convoluting parton cross sections with the proton structure functions. As seen in Fig.2,\cite{Anchordoqui:2008di} $qg \to qg$  is the most favorable channel for discovering string resonances, with an optimal balance of luminosity and color factors.
\begin{figure}[h]
\centerline{\psfig{file=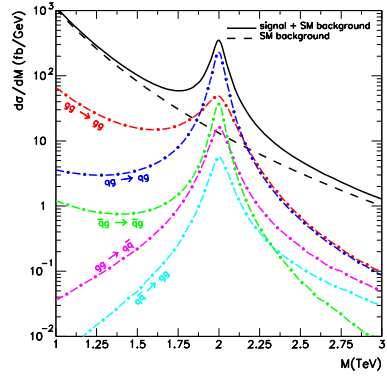,width=2.0in}}
\vspace*{8pt}
\caption{A peak in the dijet cross section $d\sigma/dM$ (units of fb/GeV) vs. M (TeV) is plotted for the case of SM QCD
background (dashed line) and (first resonance) string signal + background (solid line),
for the hypothetical (already excluded, see below) string mass value of 2 TeV and LHC $\sqrt{s}=14$ TeV. The dot-dashed
lines indicate contributions of individual channels, illustrating the dominance of $qg \to qg$.}
\end{figure}

No significant bumps in dijet mass distributions have been observed yet at the LHC in the initial runs at $\sqrt{s}=7$ TeV and $\sqrt{s}=8$ TeV. The results of  CMS searches for string resonances, based on the integrated luminosity of 4 fb$^{-1}$ at $\sqrt{s}$= 8 TeV, are shown on Fig.3.\cite{Chatrchyan:2013qha} The data are presented as the observed upper limits (at 95\% confidence level) on $\sigma\times \mathcal{B}\times A$ where $\mathcal{B}\approx 1 $ is the branching fraction of $q^*$ into $qg$ and $A\approx 0.6$ is the acceptance for the respective kinematic requirements. The cross section $\sigma$ is integrated over the mass bins determined by the uncertainty in jet energy resolution.\footnote{Even at $M= 4$ TeV, the width  of $q^*$ resonance, $\Gamma_{q^*}< 150$ GeV,\cite{Anchordoqui:2008hi} is sufficiently small to justify the narrow resonance approximation applied to CMS data.} Based on these data, CMS quote the lower bound, $M>4.78$ TeV, on the string mass.\footnote{CMS has also ruled out excited top quarks with $M_{t^*}<790$ GeV by searching in the top quark plus gluon decay channels.\cite{rs} This provides an additional bound on a class of ``warped'' string compactifications.\cite{Hassanain:2009at}}
\begin{figure}[h]
\centerline{\psfig{file=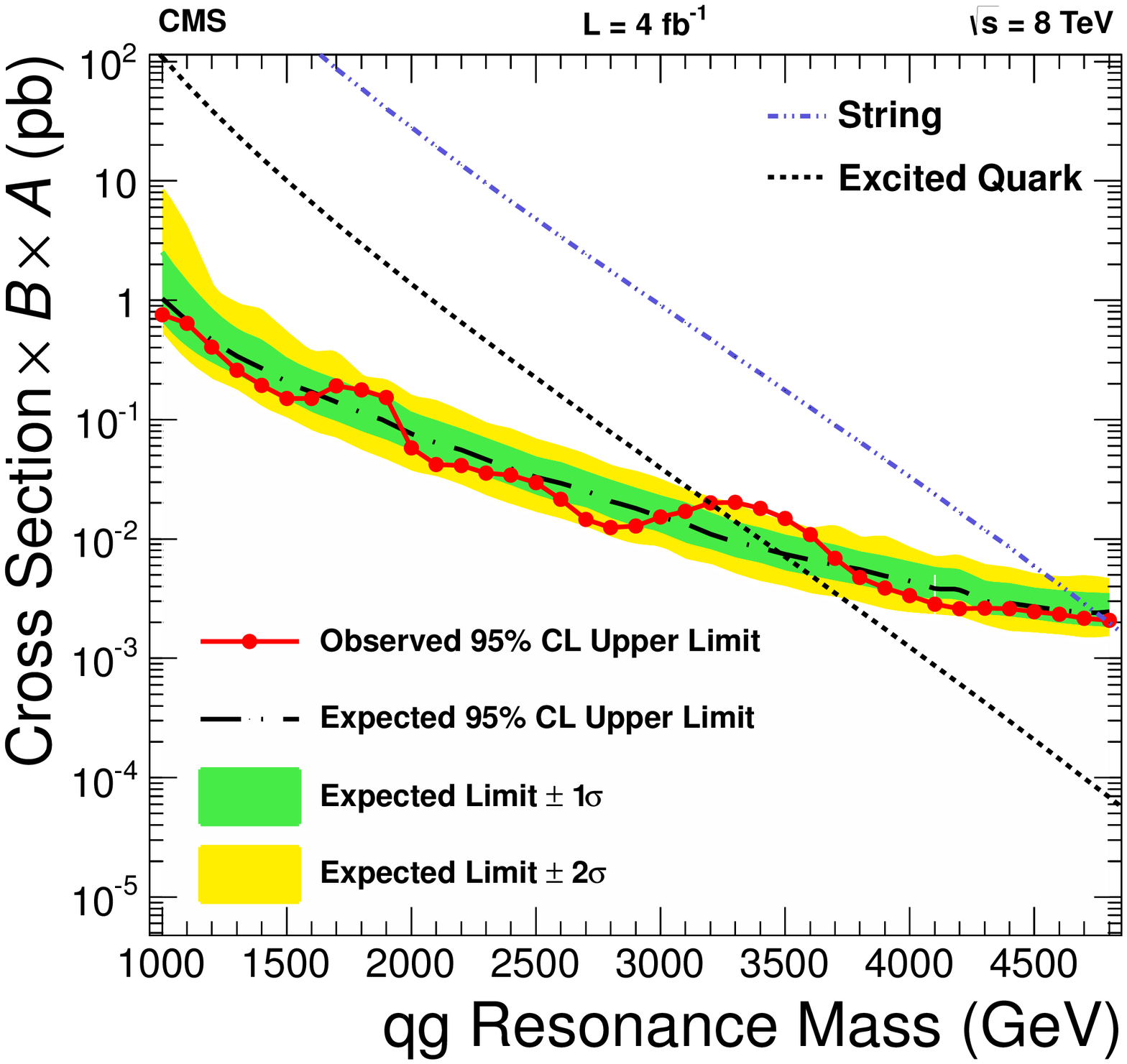,width=2.0in}}
\vspace*{8pt}
\caption{Observed upper limits on $\sigma\times \mathcal{B}\times A$ for resonances decaying into  $qg$ final state (points and solid lines). The predicted cross section, marked as ``String'', is based on the assumption of $q^*$ $j=1/2$ and $j=3/2$ resonances.}
\end{figure}

Low mass string theory will get another chance in the next run. As seen in Fig.4,\cite{Anchordoqui:2008di} string mass scales as high as 7 or even 8 TeV can be reached by LHC operating at its design center of mass energy $\sqrt{s}=14$ TeV.
\begin{figure}[h]
\centerline{\psfig{file=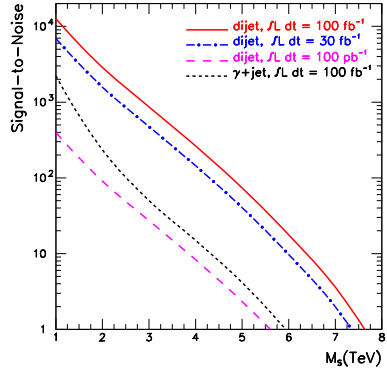,width=2.0in}}
\vspace*{8pt}
\caption{Dijet signal-to-noise ratio for string resonances at the LHC with the design $\sqrt{s}=14$ TeV, for three integrated luminosities.}
\end{figure}

\section{High (and low) string scale: extra $U(1)$ gauge bosons}

No matter what the string scale is, all D-brane models predict the existence of more neutral gauge bosons than  the standard model. The reason is that gauge groups are associated to stacks of D-branes and in this framework, the color $SU(3)$ group emerges from a stack of at least three branes which, in the minimal case, generate $U(3)=SU(3)\times U(1)$, with the extra $U(1)$ coupled to the baryon number $B$. Since $B$ is an anomalous symmetry, the corresponding $Z'$ gauge boson receives a mass via Green-Schwarz mechanism, of order $g'M<M$. In general, there are more than one extra gauge bosons. They fall into two classes:
anomalous, massive $Z'$ or non-anomalous $Z^{\prime\prime}$ with the mass generated by vacuum expectation values of an extended Higgs sector. In the latter case, there is no direct relation between the mass and the string scale $M$.

There are many other extensions of the standard model involving extra gauge bosons, so such a particle would constitute at most a ``harbinger'' of string theory. It is natural to expect that $Z'$ and/or $Z''$ couple to quarks: in this case a bump would be observed in the dijet invariant mass spectrum. Unfortunately, no extra gauge bosons have been found yet at the LHC, see Fig.5.\cite{Chatrchyan:2013qha} In this case, CMS collaboration quotes $M_{Z'}>1.62$ TeV assuming that $Z'$ couples with the same strength as $Z$.
\begin{figure}[h]
\centerline{\psfig{file=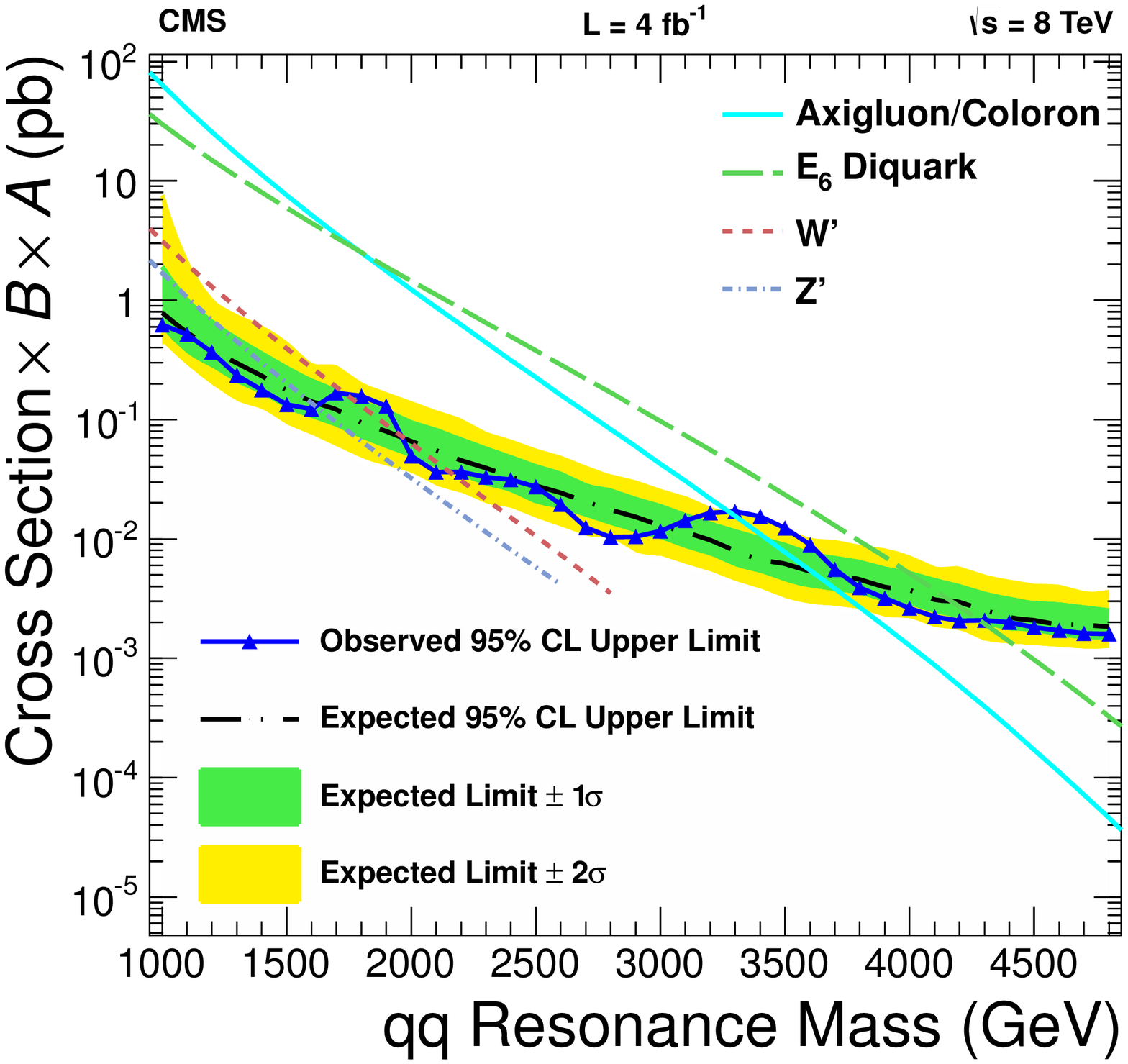,width=2.0in}}
\vspace*{8pt}
\caption{Limits on various resonances, including $Z'$,  decaying into  $q\bar q$ final state.}
\end{figure}

If the string mass is too high for the string resonances or $Z'$ to be discovered at the LHC, $Z''$ gauge bosons, with masses determined by an extended Higgs sector, still remain as a possibility. In a minimal model of that type, called SM$^{++}$,\cite{Anchordoqui:2012wt} one additional D-brane
intersects the $U(3)$ stack. The non-anomalous $Z''$ couples to the combination of $B{-}L$ ($L$ is the lepton number) and the right-handed isospin $I_R$. The strength of its $q\bar q$ coupling depends on the combination, and the existing dijet data can be interpreted in this context to establish more specific bounds on $Z''$ mass, as seen in Fig.6.\cite{Anchordoqui:2012wt} The LHC, operating at its design center of mass energy, will explore a wider range of $Z''$ masses.
\begin{figure}[tbp]
\begin{minipage}[t]{0.49\textwidth}
    \psfig{file=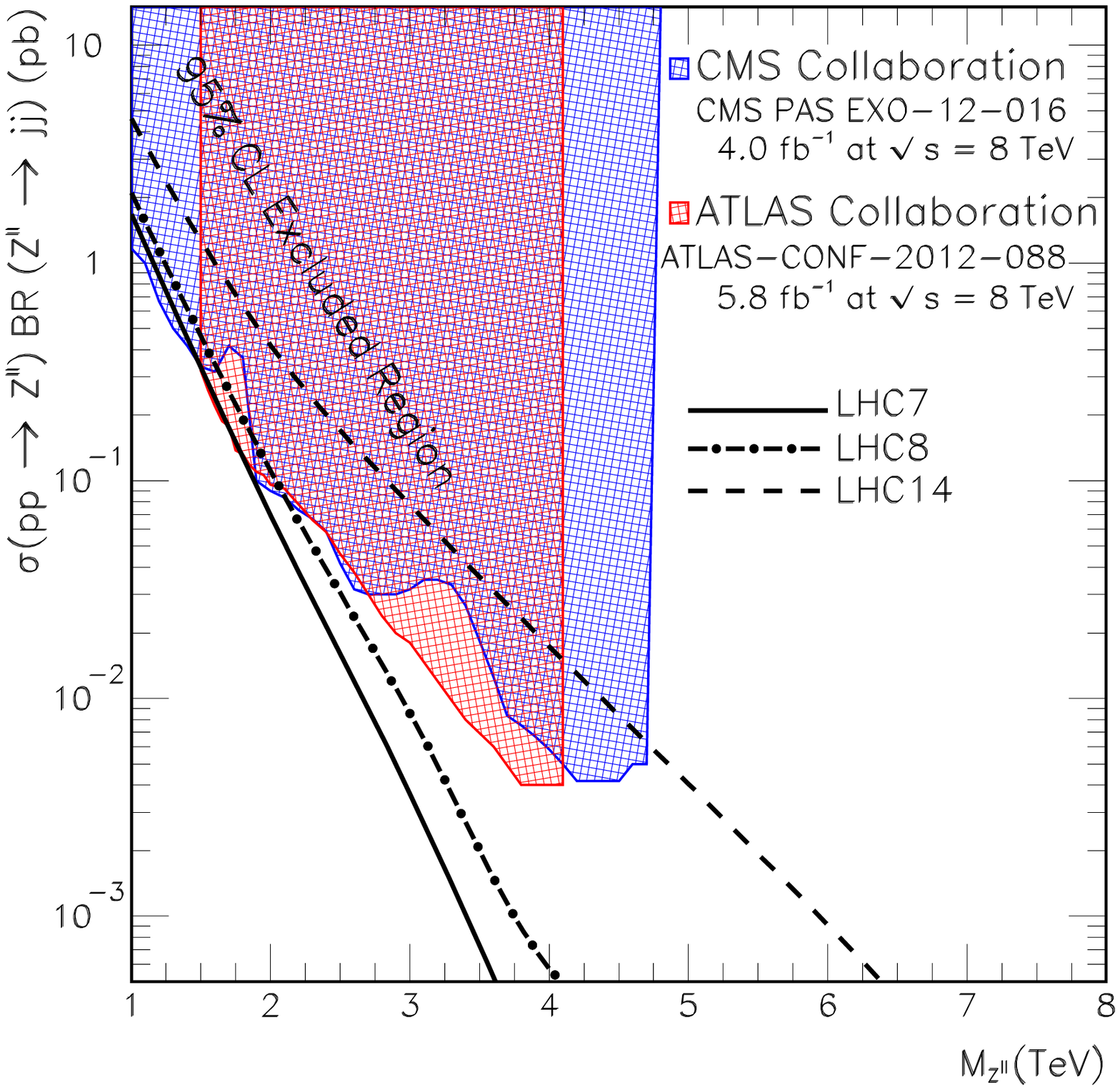,width=2.0in} \end{minipage}
  \hfill \begin{minipage}[t]{0.49\textwidth}
    \psfig{file=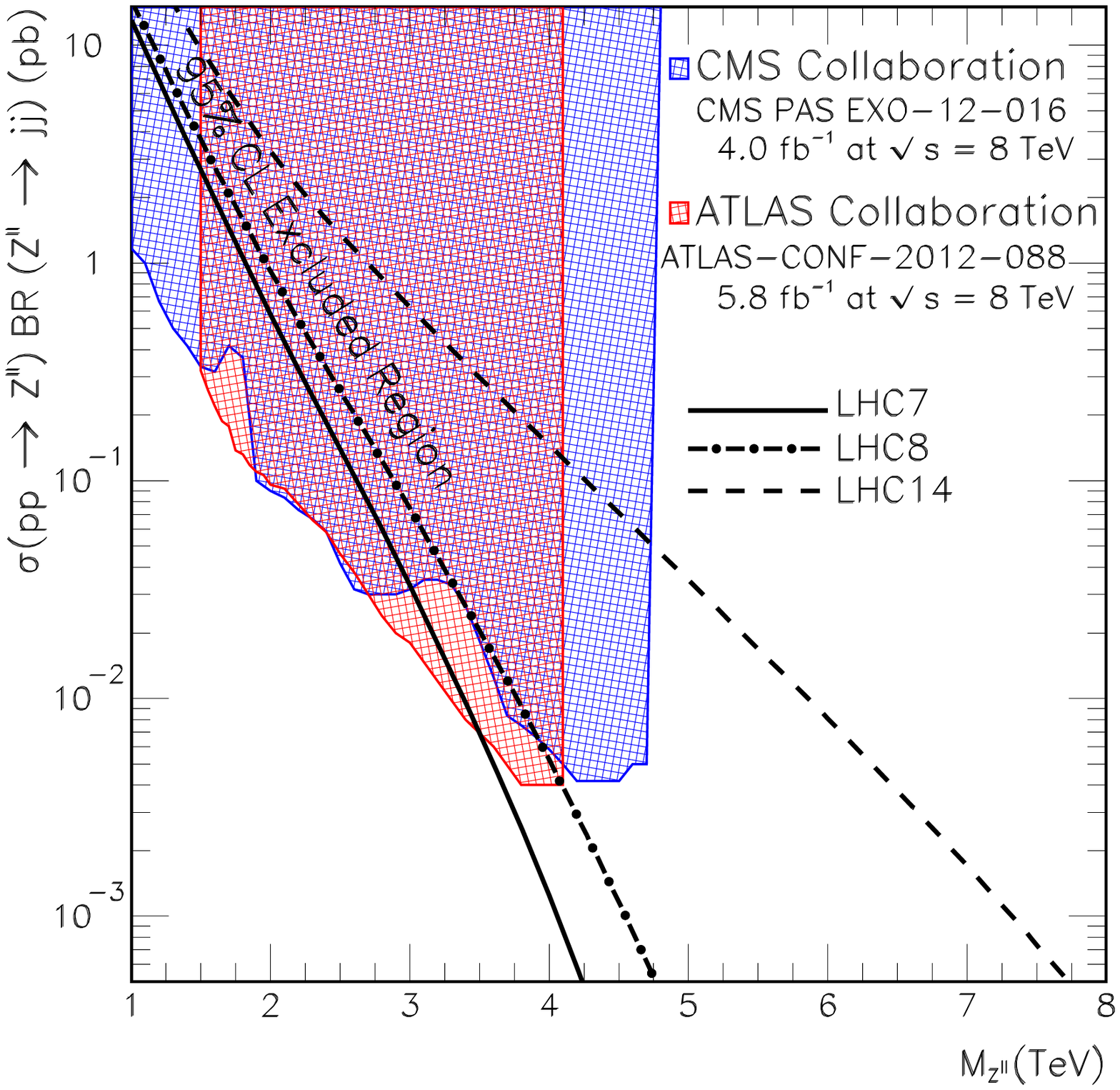,width=2.0in}\end{minipage}
  \caption{Comparison of the total cross section for the
    production of $p p \to Z'' \to jj$ with the 95\% CL upper limits
    on the production of a gauge boson decaying into two jets as
    reported by the CMS and ATLAS collaborations (corrected by
    acceptance). The case in which $Z''$ is
    mostly diagonal in $I_R$ is shown in the left panel and the case
    in which it is mostly $B-L$ in the right panel.}
\label{fig:CMS}
\end{figure}

\section{Conclusions}
No low mass string or the associated extra gauge bosons have been observed in the LHC runs at $\sqrt{s}=7$ TeV and $\sqrt{s}=8$ TeV. Unlike in many other beyond the standard model extensions, due to spectacular strength of dijet signals, low mass strings will be either discovered at the early stage of the next run or more stringent bounds will be established very soon. With the present bound $M>4.78$ TeV, the best we can hope for is the discovery of the lowest, $n=1$ string excitations of quarks and gluons because higher levels and the accompanying mini black holes will be most likely beyond the reach of LHC. Nevertheless, with such large cross sections, it should be possible to identify higher spin string excitations of gauge bosons, quarks and leptons.

\section*{Acknowledgments}
This work was partially supported by the ERC Advanced Grant ``Strings and Gravity"
(Grant.No. 32004) and by the DFG cluster of excellence ``Origin and Structure of the Universe".
In addition, this material is based in part upon work supported by the National Science Foundation under Grant No.\ PHY-0757959.  Any
opinions, findings, and conclusions or recommendations expressed in
this material are those of the authors and do not necessarily reflect
the views of the National Science Foundation.

\end{document}